\documentclass[12pt]{article}

\usepackage[T1]{fontenc}
\usepackage{latexsym}
\usepackage{physics}
\usepackage{amsmath}
\usepackage{scrextend}

\usepackage{amssymb}
\usepackage{slashed}
\usepackage{empheq}
\usepackage[english]{babel}
\usepackage{booktabs} 
\usepackage{array}
\usepackage{verbatim} 
\usepackage{xcolor}
\usepackage{soul}
\usepackage[pdfa]{hyperref}
\usepackage[]{pdfpages}
\usepackage{graphicx}
\newcommand{\bbar}{\overline}

\begin{document}
\title{Albert Einstein and the fifth dimension.\\
A new interpretation of the papers published in 1927}

\author{Giulio Peruzzi\thanks{{giulio.peruzzi@unipd.it, Department of Physics and Astronomy ``G. Galilei'', via Marzolo 8, I-35131 Padova (Italy)}} 
\and Alessio Rocci\thanks{{a\_rocci@hotmail.com, Department of Physics and Astronomy ``G. Galilei'', via Marzolo 8, I-35131 Padova (Italy)}}}

\maketitle
\begin{abstract}
In 1927 Einstein sent two brief communications to the Prussian Academy of Sciences on Kaluza's five-dimensional theory. In his Einstein biography, Abraham Pais asserted that he could not understand the reasons that pushed Einstein to communicate his work. Indeed, Einstein's paper seems to be very close to Oskar Klein's approach, published in 1926. The question seems to be yet unanswered, also in recent works. We analysed the differences between Einstein's and Klein's work and we propose a new interpretation of Einstein's approach. In 1927, he tried to use Weyl's scale invariance in order to construct a theory covering macroscopic as well as microscopic phenomena. In constructing a five-dimensional action, Einstein used in a modern way the idea of ``gauge invariance''. Furthermore, we propose an additional motivation for the fact that he did not  mention Klein's and Vladimir Fock's contributions in his first communication and we show further evidences which would confirm also Paul Halpern's recent proposal. As a consequence of our interpretation of the role of conformal transformations, we suggest that in 1927 Einstein did not consider the fifth dimension as real and that his approach, unlike Klein's attempt, cannot be considered inconsistent. 
\end{abstract}

\tableofcontents

\section{Prologue: inconsistent!}\label{prologue}
In his first paper on the five-dimensional Universe \cite{Klein1}, in order to unify gravitational, electromagnetic and Schr\"odinger's description of the electron's motion, Oskar Klein introduced a five-dimensional space-time\footnote{See \cite{Klein1-trad} for an English translation.}. In Klein's model, Einstein's General Relativity (GR) and Maxwell's electromagnetism (EM) emerged as part of a five-dimensional GR-like theory, while the quantum behaviour of the electron was represented by a ``light-like'' five-dimensional wave equation. Klein showed that, using suitable Ansatz, the five-dimensional Einstein equations, obtained using the five-dimensional curvature scalar\footnote{In our paper, the tilde-signed  quantities will refer to five-dimensional space-time.} $\displaystyle \tilde{R} $ as Lagrangian density, are equivalent to the four-dimensional Einstein equations coupled with Maxwell's equations. Furthermore, using the same assumptions, he showed that in the geometrical optics approximation the rays of the five-dimensional light wave, i.e. the five-dimensional null geodesics, are equivalent, via a suitable projection, to the four-dimensional Lorentz equation for a charged particle moving both in a gravitational and in a electromagnetic field. Let us see how Klein presented and used his assumptions. 

As first assumption, Klein imposed the so-called \emph{cylinder condition}, namely\footnote{We adopted the following conventions. Barred indices refer to the five-dimensional coordinates, $ \bar\mu = 0, 1, 2, 3, 5$, where the zero component corresponds to a time-like dimension. We use the mostly-plus signature, i.e. $\eta_{\bar\mu\bar\nu}=diag(-1,+1,+1,+1,+1) $. The unbarred Greek indices correspond to the usual four-dimensional space-time, $ \mu = 0,1,2,3 $, and Latin indices refer to the three-dimensional spatial coordinates, $ i = 1,2,3 $.} $\partial_5\gamma_{\bar{\mu}\bar{\nu}} = 0 $, where $ \gamma_{\bar{\mu}\bar{\nu}} $  is the the five-dimensional metric\footnote{This condition was introduced for the first time by Theodore Kaluza \cite{Kaluza}, as we shall review in section \ref{Kaluza}. See \cite{Kaluza-trad} for an English translation.}. As a consequence, using modern language, there is a residual four-dimensional general coordinate invariance and an invariance associated with the transformations of the fifth coordinate. Indeed, the cylinder condition means that the five-dimensional space-time admits a Killing vector, i.e. a preferred direction in five dimensions.

As second assumption, Klein emphasized that the scalar quantity $ \gamma_{55} $ is invariant under the transformation laws presented and that an admissible choice would be $\gamma_{55} = \text{constant} $. This condition was dubbed \textit{sharpened cylinder condition} by Albert Einstein\footnote{We adopted the English translation introduced in \cite{Einstein-papers15bis} of Einstein's definition.} in 1927. In order to motivate his choice from a physical point of view, Klein suggested that the sharpened cylinder condition can be considered as a mere convention, once it is conjectured that `only the ratios of $ \gamma_{\bar{\mu}\bar{\nu}} $ have physical meaning.' (\cite{Klein1}; p. 896).

Thanks to these two postulates, the five-dimensional quantities  $ \gamma_{\bar{\mu}\bar{\nu}} $ are reduced to the four-dimensional set of fourteen variables, namely $ g_{\mu\nu} $ and $ A_\mu $, i.e. the gravitational and the electromagnetic potentials. Then, Klein emphasized that a unified theory would address the problem of the field equations for all of the $ \gamma_{\bar{\mu}\bar{\nu}} $ `from which the field equations for $ g_{\mu\nu} $ and $ A_\mu $ in the ordinary theory of relativity emerge in a suitable approximation.' (\cite{Klein1}; p. 897), but he explicitly decided to avoid the discussion of this `difficult problem'(\cite{Klein1}; p. 897). Klein proposed $ \tilde{R}\sqrt{-\gamma} $ as Lagrangian for constructing a five-dimensional action, where $ \sqrt{-\gamma} $ is the determinant of the metric, then he inserted both hypotheses into the five-dimensional action. Therefore, Klein implicitly suggested that $ \tilde{R}\sqrt{-\gamma} $ would be a sort of effective action obtained by five-dimensional constrained Lagrangian. He motivated his choice by showing that the five-dimensional field equations obtained by varying $ \gamma_{5\mu} $ are equivalent to the system formed by Einstein's and Maxwell's four-dimensional equations.

The constancy of $\gamma_{55} $ is sometimes claimed as responsible for an inconsistency of Klein's model \cite{KK-math2}, \cite{KK-math},  \cite{KK-rev}, \cite{Raife2000}. The inconsistency emerges if we consider the full five-dimensional field equations, treating $\gamma_{55} $ as a function of the space-time coordinates, and the meaning of the extra-dimension. Indeed, after having defined the electromagnetic potentials as follows, namely
\begin{equation}\label{def-A}
	\kappa A_\mu = \frac{\gamma_{5\mu}}{\gamma_{55}}
\end{equation} 
 where\footnote{$ G $ is the four-dimensional Newton constant and $ c $ is the speed of light.} $ \kappa = \frac{8\pi G}{c^4} $, the $55-$component of the five-dimensional Einstein equations reads \cite{Raife2000}: 
\begin{equation}\label{55-eq}
	\square \sqrt{\gamma_{55}} = \frac{\kappa^{2}}{4} \left( \sqrt{\gamma_{55}}\right)^3 F_{\alpha\beta}F^{\alpha\beta} \; ,
\end{equation}
where the four-dimensional operator $ \square $, when acting on the scalar function $ \gamma_{55}(x) $ is defined by $ \square \gamma_{55} =g^{\mu\nu}\nabla_{\mu}\partial_{\nu}\gamma_{55}$ for a curved four-dimensional space-time, and $ \nabla_{\mu} $, and $F_{\alpha\beta}=\partial_{\alpha}A_{\beta}-\partial_{\beta}A_{\alpha}$ represent the covariant derivative and the Maxwell's antisymmetric tensor respectively. By interpreting the fifth dimension as a physical extra-dimension, the dimensionally reduced field equations must be consistent with the five-dimensional equations (\cite{KK-rev2}; p. 8-9). The sharpened cylinder condition is inconsistent with the five-dimensional field equations, because it would imply the too restrictive condition $F_{\alpha\beta}F^{\alpha\beta} = 0 $, which means that the moduli of the electric and the magnetic field should be proportional to each other. As emphasized in \cite{Raife2000}, the inconsistency is implied by the fact that Klein gave a physical meaning to the fifth dimension, in order to incorporate Schr\"odinger's wave mechanics in his unified model\footnote{In \cite{Raife2000} it is also emphasized that the inconsistency was pointed out for the first time by Pascual Jordan \cite{J1} and Yves Thiry \cite{T1} in 1947 and in 1948 respectively.} in \cite{Klein1} and to justify the quantization of the electric charge in \cite{Klein2}.

Soon after Klein's work, in February 1927, Einstein published his first papers on the five-dimensional Universe \cite{E27a} \cite{E27b}, which have given room to unanswered questions. Einstein had discussed the five-dimensional approach with Theodore Kaluza already in 1919, who tried to unify the electromagnetic and the gravitational forces. Kaluza published his paper in 1921 \cite{Kaluza} and he did not introduce the sharpened cylinder condition. Abraham Pais pointed out how Einstein's attitude toward the five-dimensional approach changed between the beginning and the end of August 1926. Then, referring to Einstein's 1927 papers, he emphasized: `I should explain why these papers are a
mystery to me. [...] What does
puzzle me is a note added to the second paper [...] I fail to understand why he published his two notes in the first place.' (\cite{Pais}; p. 333). Einstein's papers are two brief communications. As he himself emphasized in a note added in proof `[...] the findings [...] are not new. The entire content is found in the paper by O. Klein \cite{Klein1}. Compare furthermore Foch's paper \cite{Fock}.' (\cite{Einstein-papers15bis}; p. 478). This is the note Pais referred to. Why did Einstein published these two communications? This is the first unanswered question regarding this Einstein's work, even if it's worth noting that recently Paul Halpern offered a possible explanation by discussing why Einstein did not quote Klein's paper in his first communication \cite{Halpern}. The other questions are connected with the first one. Einstein had discussed the role of $ \gamma_{55} $ with Kaluza and they analyzed the effect of a physical fifth dimension on four-dimensional particle's dynamics, as it emerges clearly from a footnote of Kaluza's paper (\cite{Kaluza}; p. 970). Was Einstein aware of the inconsistency of the sharpened cylinder condition we described above? Furthermore, what was Einstein's attitude toward the character of the fifth dimension in 1927? Even if he claimed the results were not new with respect to Klein's work, is there any difference between Einstein's and Klein's approach? Finally, unlike Klein, Einstein did not consider Schr\"odinger's wave mechanics. Did Einstein want to unify electromagnetic and gravitational forces only, or did he want to incorporate microscopic phenomena as well?

This paper aims at addressing these questions. Even if many authors have analyzed Kaluza's and Klein's approach \cite{Raife2000} \cite{DawningGauge} \cite{Goenner} \cite{VanDongen} \cite{Vizgin}, Einstein's communications are less known and only short comments can be found (\cite{Halpern}, \cite{Goenner}, \cite{VanDongen}, \cite{Vizgin}, \cite{Sauer} and references therein). Furthermore, Tilman Sauers explicitly emphasized: `We still lack fine-grained historical investigation of [Einstein's] later work [...], that is, investigations that would discuss his endeavors with technical understanding from a historical
point of view [...]' (\cite{Sauer}; p. 282). Therefore, with the help of a recent translation contained in the last published volume of the Collected Papers of Albert Einstein (CPAE) project \cite{Einstein-papers15bis}, in section \ref{Ein-27} we start with a closer inspection of Einstein's communications. We shall emphasize that Einstein's approach was different with respect to Klein's attempt. We propose a new interpretation of Einstein's papers, which would offer another explanation for Pais's perplexity: first, we argue that he wanted to build up explicitly what we call, in modern language, a gauge theory; second, we shall point out the role of the conformal transformations. In order to discuss Einstein's attitude toward the fifth dimension and therefore the consistency of his model, in section \ref{Kaluza} we shall reconsider Einstein-Kaluza correspondence and the role of the fifth dimension in Einstein's work until 1927. We argue that Einstein was aware of the inconsistency of the sharpened cylinder condition with a physical extra-dimension and that its formulation in coordinate-free form was one of the goal that forced him to publish the two communications in 1927. We propose the following interpretation of Einstein's approach. In 1927, Einstein was convinced that the dynamics of fields and particles should not be affected by the fifth dimension. In order to achieve this goal, Einstein discussed the role of conformal transformations, which would give the fifth dimension a mere mathematical meaning. Therefore, in section \ref{Weyl}, we shall reconsider Einstein's attitude towards Hermann Weyl's ideas. As a result, we propose to interpret his introduction of the ``conformal invariance'' of the four-dimensional physics like an attempt to incorporate the microscopic phenomena, without the help of Schr\"odinger's wave mechanics.
In section \ref{epilogue} we summarize our point of view and  we conclude that, by accepting our interpretation, Einstein's approach can be regarded as consistent, even if he introduced the sharpened cylinder condition.

\section{Einstein's communications}\label{Ein-27}
\subsection{The first communication}\label{Ein-27a}
In the introduction of his first communication, Einstein considered the two main attempts to unify gravitational and electromagnetic forces. Indeed, he referred to Weyl's and Arthur Eddington's attempt on one side, and he quoted Kaluza's paper \cite{Kaluza} on the other side. The aim of the first two authors, in Einstein's words, was `to bring together Gravitation and Electricity into a unifying framework [...] through a generalization of the Riemannian geometry' (\cite{E27a}; p. 23). Kaluza's attempt, instead, `maintained Riemannian metric, but introduced a five-dimensional space-time, which could be reduced to some extent to a four-dimensional space-time through the >>cylinder condition<<.' (\cite{E27a}; p. 23). Einstein's purpose in his first communication was `to draw attention to a disregarded point of view, which is \textit{essential} for the Kaluza's theory.' [our emphasis] (\cite{E27a}; p. 23). This extremely important ingredient of Kaluza's theory is the cylinder condition. Unlike Klein, as emphasized in \cite{Einstein-papers15}, Einstein gave an explicitly coordinate-independent formulations of the cylinder condition. Indeed, he reformulated the cylinder condition in terms of Killing vectors. We point out that this is the first time that Killing equations for the isometries of the metric appeared in a physics paper. After having introduced the five-dimensional space-time, Einstein associated the cylindrical shape of the space-time manifold with the existence of a normalized displacement-vector field $ \xi^\mu $, which generates isometry transformations of the metric. Therefore, the $ \xi^\mu $ vector must satisfy a particular equation, which is known nowadays as the Killing equation, because it was first introduced by Wilhelm K. J. Killing (\cite{Killing}; p. 167). Einstein wrote it in the following form (\cite{E27a}; p. 23, eq.2), namely:
\begin{equation}\label{Cyl-Kill}
\xi^{\bar{\beta}}\partial_{\bar{\beta}}\gamma_{\bar{\mu}\bar{\nu}} + \gamma_{\bar{\beta}\bar{\nu}}\partial_{\bar{\mu}}\xi^{\bar{\beta}} + \gamma_{\bar{\beta}\bar{\mu}}\partial_{\bar{\nu}}\xi^{\bar{\beta}}= 0\; .
\end{equation}
Then, using the freedom to choose the coordinate-system, Einstein pointed out that the cylinder condition assumes the form stated by Kaluza, i.e. $\partial_5\gamma_{\bar{\mu}\bar{\nu}} = 0 $, if the invariant direction points to the fifth coordinate\footnote{Einstein explicitly chose $ \xi^5 $ as the only non-vanishing component. Hence, the Killing vector assumes the following form: $ \xi^{\bar{\beta}}=(0,0,0,0,1) $ and, inserting it into eq. (\ref{Cyl-Kill}), Einstein obtained Kaluza's condition.}. It is worth noting that Einstein underlined, in a footnote, the role of equation (\ref{Cyl-Kill}), which states explicitly that Kaluza's cylinder condition can be recast in a \textit{manifestly covariant form} (\cite{E27a}; p. 23). Hence, we can infer that for Einstein was an important fact to express the cylinder condition in a manifestly covariant form. The importance of eq. (\ref{Cyl-Kill}) is witnessed also by a statement of the original manuscript (\cite{Einstein-papers15}; p. 720, note [7]), where Einstein emphasized that Kaluza tried to combine the cylinder condition and the request of full covariance in five-dimensions in a unnatural way and that with eq. (\ref{Cyl-Kill}) he found a more natural way for introducing the cylinder condition.

Why did Einstein consider as fundamental to understand if the cylinder condition can be recast in a manifestly covariant form? How long did Einstein struggle to find it? In order to answer these questions, we shall reconsider, in section \ref{Kaluza}, the Kaluza-Einstein correspondence at the time when both authors started to discuss it and we shall see that Einstein struggled with the cylinder condition from the very beginning (1919). Because of the emphasis that Einstein gave in the paper to his finding of a covariant formulation and because of his long quest for a coordinate-free form, we propose to identify this ingredient as one of the motivations that pushed him to publish his communications. 

After having considered the cylinder condition, Einstein was concerned with the role of $\gamma_{55}  $ itself. As we said in section \ref{prologue}, Klein decided to set $\gamma_{55}=\text{constant}  $ after having presented the invariance group of transformations implied by the cylinder condition. Unlike Klein, Einstein investigated the geometrical meaning of this condition: this is another reason for stating that Einstein's approach was differenf from Klein's. Einstein noticed that even if the non-zero component of the Killing vector is constant, on a curved manifold, its modulus is not necessarily constant. Hence, Einstein noted that setting the modulus of the Killing vector to be constant over the whole five-dimensional space-time implied the constancy of $\gamma_{55}  $. Indeed, the squared modulus of the Killing vector $ \xi^2 $ can be rewritten as follows: $ \xi^2 = \gamma_{\bar{\mu}\bar{\nu}} \xi^{\bar{\mu}}\xi^{\bar{\nu}} =\gamma_{55} \xi^5\xi^5 =\gamma_{55}$, therefore $ \xi^2 = \text{constant}$ is equivalent to  $\gamma_{55} =constant $. Einstein called this last constraint \textit{sharpened cylinder condition}. As Einstein promised, he analyzed Kaluza's theory from a different perspective. Indeed, he adopted a more geometrical point of view. As we shall see in the following, the coordinate transformations presented also by Klein emerged, in Einstein's paper, as a consequence of the geometry of the five-dimensional space-time. At this stage, Einstein observed that once assumed $\gamma_{55} = constant $, either $\gamma_{55} = 1 $ or $\gamma_{55} = -1 $ are equally acceptable choices. He considered the first option, promising to discuss the other option on another occasion. He will come back to this point in the second communication, by discussing the variational principle which generates the field equations.  

In 1927 Einstein did not yet realized that a manifold which admits a Killing vector with constant modulus is equivalent to a manifold where the Killing trajectories are geodesic lines. As far as we know, Heinrich Mandel would point out this important fact for the first time two years later (\cite{Mandel3}; p. 564).  As Einstein himself would write at the end of his second communication, Mandel attracted Einstein's attention on the priority of Fock's and Klein's papers. Indeed, Mandel had started to investigate the five-dimensional Universe with non-constant $ \gamma_{55} $ in \cite{Mandel1}, where he also discussed the physical meaning of the extra-dimension. In \cite{Mandel1} the author did not consider the role of Killing equations, while in August 1926, Mandel would took Einstein's point of view. Indeed, in his following paper, after having quoted Einstein's communications, Mandel used for the first time the expression ``Killing equations'' (\cite{Mandel2}; p. 290) and, following Einstein, noticed that the cylinder condition takes Kaluza-Klein's form for a special coordinate system. Two years later, Mandel would realize the connection between the cylinder condition and the geodesic character of the Killing trajectories. It is worth noting that Einstein would introduce the same idea only eleven years later (\cite{E38}; p. 684) without quoting Mandel's paper. This fact, points out the importance of this two communications for understanding how Einstein's ideas evolved. 

After having analyzed the geometric meaning of the sharpened cylinder condition, Einstein inserted the two hypotheses into the five-dimensional line element and showed how gravitational as well as electromagnetic potentials emerge from the five-dimensional metric tensor, pointing out how Kaluza's theory should unify both forces in a natural way. Subsequently, Einstein considered only the cylinder condition in the `adapted coordinate system' (\cite{E27a}; p. 24), where the invariant direction has been set along to the fifth coordinate. Like Klein, using modern language, Einstein underlined that the cylinder condition implied that Kaluza's theory invariance group can be written as a product of the four-dimensional diffeomorphism group and a one-dimensional group. In order to investigate the role of this residual symmetry, Einstein relaxed explicitly the sharpened cylinder condition and called the one-dimensional group that leaves invariant the four-dimensional metric `the $ x^5\text{-transformations} $' (\cite{E27a}; p. 24-25), namely\footnote{In Einstein's original paper he used Latin indices for the four-dimensional continuum.} (\cite{E27a}; p. 25):
\begin{eqnarray}
x^{\mu} &=& \bbar{x^{\mu}}    \label{four-dim-diffeo}\nonumber\\
x^5 &=& \bbar{x^5} + \psi\left(\, \bbar{x^0},\, \bbar{x^1},\, \bbar{x^2},\, \bbar{x^3}\,\right) \; ,\label{x5-trasf}
\end{eqnarray}    
where a bar over a quantity, in this section, will indicate the same object but in the transformed coordinate system and $ \psi $ is an arbitrary function of the four-dimensional coordinates only.
Then, Einstein identified his ``adapted coordinate system'' with a specific four-dimensional hypersurface embedded into the five-dimensional space-time manifold.

Unlike Klein, Einstein wrote explicitly how the components of the five-dimensional metric tensor would transform under the action of the $ x^5-$transformations, without the assumption of the sharpened cylinder condition (\cite{E27a}; p. 25):
\begin{eqnarray}
\gamma_{\mu\nu} &=& \bbar{\gamma_{\mu\nu}}+\frac{\partial \psi}{\partial\bbar{x^\mu}}\bbar{\gamma_{5\nu}}+\frac{\partial \psi}{\partial\bbar{x^\nu}}\bbar{\gamma_{5\mu}}+\frac{\partial \psi}{\partial\bbar{x^\mu}}\frac{\partial \psi}{\partial\bbar{x^\nu}}\bbar{\gamma_{55}}\label{trasf-gamma1}\\
\gamma_{5\nu} &=& \bbar{\gamma_{5\nu}}+\frac{\partial \psi}{\partial\bbar{x^\nu}}\bbar{\gamma_{55}}\label{gauge}\\
\gamma_{55} &=& \bbar{\gamma_{55}}\label{trasf-gamma2}\; .
\end{eqnarray}
But applying the sharpened cylinder conditions to  eq. (\ref{gauge}), Einstein recognized the gauge transformations for the electromagnetic potentials. After having set $ \gamma_{55}=1 $,
like Kaluza, Einstein defined the electromagnetic potentials and the four-dimensional metric tensor as follows\footnote{We remember that $ \kappa = 1 $.}:
\begin{eqnarray}
A_\mu &=&  \gamma_{5\mu}\label{var1}\\
g_{\mu\nu} &=& \gamma_{\mu\nu}-\gamma_{5\mu}\gamma_{5\nu}\label{var2} \; ,
\end{eqnarray} 
and he pointed out the invariance of the four-dimensional metric tensor under the action of the $ x^5\text{-transformations} $ and the connection between $ x^5\text{-transformations} $. 

At the end of his first communication, Einstein made explicit his strategy for constructing the \textit{physical} five-dimensional theory. Einstein's purpose was to construct a Lagrangian density, in order to obtain the usual gravitational and electromagnetic field equations. But in this first communication, Einstein did not tackle the whole problem and presented only an argument concerning electromagnetic phenomena. Einstein proposed a specific argument, which permitted him also to justify his conviction that the electromagnetic potentials have no physical meaning\footnote{The physical reality of the electromagnetic potentials would be recognized only after the Aharonov-Bohm effect \cite{Aharonov-Bohm1} \cite{Aharonov-Bohm2}.} and that he started to justify already in 1921\footnote{See section \ref{KE-correspondence}.}. Indeed, Einstein observed that if the Lagrangian density `is also supposed to be invariant with respect to the $ x^5- $transformations [...], then this invariant may only contain the $ \gamma_{5\mu} $'s in the combinations $ \displaystyle F_{\mu\nu} = \partial_{\mu} \gamma_{5\nu} -\partial_{\nu} \gamma_{5\mu}  $.' (\cite{E27a}; p. 25), i.e. it should contain only the electromagnetic antisymmetric tensor, namely $ \displaystyle F_{\mu\nu} = \partial_{\mu} A_\nu -\partial_{\nu} A_\mu  $. Finally, Einstein pointed out again  the importance of equation (\ref{gauge}). This result would remain valid also without imposing the sharpened cylinder condition, as he will emphasize in the second communication, by defining $ \displaystyle A_\mu =  \frac{\gamma_{5\mu}}{\gamma_{55}} $. This fact, which is connected with the Ansatz that the space-time manifold has a symmetry group $ G $, is well known nowadays. These features have been implemented in modern Kaluza-Klein theories, where the massless states include Yang-Mills gauge fields with gauge group $ G $ (\cite{KK-rev2}; p. 15). Also Klein emphasized the analogy between the transformation law induced by the $ x^5 $-transformations and the usual gauge freedom of the electromagnetic potentials in Maxwell's theory. But, Einstein's suggested to use this postulate in order to construct an action principle. This approach resemble the procedure we apply today to construct a gauge theory. In this context, the group of the $ x^5\text{-transformations} $ played the role of a gauge group. Furthermore, in his second communication, Einstein extended his gauge approach, in order to incorporate microscopic phenomena.

Are we allowed to define Einstein's procedure as a gauge approach, in a modern sense? It is worth remembering that nowadays the isometry group is regarded like an external symmetry group, while a gauge group is usually called an internal symmetry group. This fact is more than an analogy and this is `the whole beauty of Kaluza-Klein theories' (\cite{KK-rev2}; p. 15), but did Einstein consider the $ x^5 $-transformation as a gauge transformation? Like John Norton emphasized, after Einstein's elaboration the so called ``hole argument'' and after his reply to Erich Kretschmann's objection, he was aware of the ``passive'' as well of the ``active'' reading of the general covariance. Passive general covariance means that `if we have some system of fields, we can change our space-time coordinate system as we please and the new descriptions of the fields in the new coordinates system will still solve the theory's equations.' (\cite{Norton}; p. 113). Active general covariance `licenses the generation of many solutions of the equations of the theory in the same coordinate system' (\cite{Norton}; p. 113) and the new fields are mathematically but not physically distinct fields, like the ``hole argument'' proves (\cite{Norton}; p. 114). In this sense, general covariance can be interpreted as a gauge freedom. Indeed, Einstein proposed the same interpretation for his $ x^5 $-transformation, when he connected it with the transformation laws of the electromagnetic potentials. Furthermore, Einstein connected explicitly the two points of view: when he presented Kaluza's cylinder condition as emerging from the choice of an adapted coordinate system he used the passive point of view, when Einstein emphasized the necessity of constructing invariant objects, he was proposing an active point of view's interpretation of the $ x^5 $-transformations and hence of the resulting one-dimensional group. Therefore Einstein's procedure was a primitive form of gauge approach. As we shall see in the next section, he will extend his proposal also to the conformal group. 

Before proceeding with the second communication, Einstein pointed out that the field content of the theory would change without the sharpened cylinder condition. Indeed, a non-constant $ \gamma_{55} $ would imply in addition the presence of a scalar and of a symmetric tensor field\footnote{Kaluza made a similar observation.}.

\subsection{The second communication}
The second communication opens with the arguments that would lead to the Lagrangian density for the field equations of the full theory. Einstein started by relaxing the sharpened cylinder condition. He pointed out that the following three quantities are invariant with respect to the $ x^5- $transformations, namely (\cite{E27b}; p. 26):
\begin{equation}\label{invarianti}
\frac{\gamma_{\mu\nu}}{\gamma_{55}}-\frac{\gamma_{5\mu}}{\gamma_{55}}\frac{\gamma_{5\nu}}{\gamma_{55}}\; ;\quad\quad \frac{\partial}{\partial x^\mu}\left( \frac{\gamma_{5\nu}}{\gamma_{55}}\right) -\frac{\partial}{\partial x^\nu}\left( \frac{\gamma_{5\mu}}{\gamma_{55}}\right) \; ; \quad\quad \gamma_{55}
\end{equation} 
where $ \gamma_{55} $ is a function of the four-dimensional coordinates. The invariance can be checked directly by using equations  (\ref{trasf-gamma1}), (\ref{gauge}) and (\ref{trasf-gamma2}). Einstein's first expression of eq. (\ref{invarianti}) is different from any quantity presented by Klein. Indeed, also Goenner emphasized that after having presented what Einstein called $ x^5- $transformations, Klein `did not comment on the fact that [...] further invariants are available for a Lagrangian' (\cite{Goenner}; p. 45). Then, Einstein underlined again that the Lagrangian density should be constructed by using only the combinations which appear in eq. (\ref{invarianti}).

Even if Einstein never stated explicitly that the first two invariants in eq. (\ref{invarianti}) should correspond to the four-dimensional space-time metric tensor and to the electromagnetic potentials, it is tempting to identify them with $ g_{\mu\nu} $ and $ F_{\mu\nu} $ respectively, because they coincide with the definitions adopted by Einstein when $ \gamma_{55} = 1 $. Therefore, Einstein's definition for the four-dimensional metric is different from Klein's. In order to understand this point, let us consider the most general parametrization paying attention to the underlying symmetries of the theory. Let $ \alpha $ and $ \beta $ be two arbitrary real constants. The five-dimensional line element reads:
\begin{equation}
d\sigma^2 = \gamma_{\bar{\mu}\bar{\nu}}dx^{\bar{\mu}}dx^{\bar{\nu}}= \gamma_{\mu\nu}dx^\mu dx^\nu +2\gamma_{5\mu}dx^\mu dx^5 +\gamma_{55}(dx^5)^2\; .
\end{equation}    
We define, for convenience, $ \gamma_{55} = e^{2\beta\phi} $, where $ \phi $ is a scalar field nowadays known as the dilaton field. After completing the square, the quantity $ \displaystyle \gamma_{\mu\nu}-\frac{\gamma_{5\mu}\gamma_{5\nu}}{\gamma_{55}} $ can be identified with $ e^{2\alpha\phi}g_{\mu\nu} $ and the five-dimensional line element reads\footnote{Following Einstein we posed $ \kappa = 1 $.}:
\begin{equation}\label{general-metric}
d\sigma^2 = e^{2\alpha\phi}g_{\mu\nu}dx^\mu dx^\nu + e^{2\beta\phi}d\theta^2\; ,
\end{equation}  
where $ \displaystyle d\theta =dx^5 + A_\mu dx^\mu $ and $ \displaystyle A_\mu = \frac{\gamma_{5\mu}}{\gamma_{55}} $. Different parametrizations can be obtained by setting the values for $ \alpha $ and $ \beta $, which cannot be both zero. In this scenario, Klein's model can be obtained by choosing $ \alpha =0 $ by defining $\displaystyle g_{\mu\nu}^K = \gamma_{\mu\nu}-\frac{\gamma_{5\mu}\gamma_{5\nu}}{\gamma_{55}} $, while Einstein's one by choosing $ \alpha = \beta $, which gives  $\displaystyle g_{\mu\nu}^E = \frac{\gamma_{\mu\nu}}{\gamma_{55}}-\frac{\gamma_{5\mu}}{\gamma_{55}}\frac{\gamma_{5\nu}}{\gamma_{55}} $. The two four-dimensional metrics, $ g_{\mu\nu}^K $ and $ g_{\mu\nu}^E $, are related by a conformal transformation of the four-dimensional metric, because $ g_{\mu\nu}^K = \Omega^2\left( x\right) g_{\mu\nu}^E $, where $ \Omega^2\left( x\right) $ is a scalar function, once we set $ \Omega^2 = \gamma_{55} $. The two four-dimensional metrics coincide only when $ \alpha=\beta=0 $, or equivalently when $ \phi = 0 $, i.e. $ \gamma_{55} = 1 $. It is worth noting that given two arbitrary values of $ \alpha $ and $ \beta $, with $ \alpha \neq\beta $, there exists a conformal transformation of the five-dimensional metric, $ {\gamma}'_{\bar{\mu}\bar{\nu}} = \Omega^2 \gamma_{\bar{\mu}\bar{\nu}} $, that permits to write the line element in a Klein-like form, i.e. $ \displaystyle d\sigma^2 = ds^2 + e^{2\gamma\phi}d\theta^2 $. But Einstein's choice is more subtle, because only when $ \alpha = \beta $, any arbitrary conformal transformation of the five-dimensional metric would leave invariant both $  A_\mu $ and the four-dimensional metric $ g_{\mu\nu}^E $, which are defined by using only the ratios of the five-dimensional metric. Indeed, a conformal transformation 
$\displaystyle {\gamma}'_{\bar{\mu}\bar{\nu}} = e^{2\beta\pi} \gamma_{\bar{\mu}\bar{\nu}}= \Omega^2 \gamma_{\bar{\mu}\bar{\nu}}  $ gives, for the five-dimensional line-element:
$$\displaystyle d(\sigma')^2= e^{2\beta\phi' }(ds')^2+e^{2\beta\phi'}d(\theta')^2 =e^{2\beta\pi}d\sigma^2=e^{2\beta(\phi+\pi )}ds^2+e^{2\beta(\phi+\pi )}d\theta^2 \; ,   $$
which implies $ (ds')^2 =ds^2 $ and $ d(\theta')^2 = d\theta^2 $, because $ \phi' = \phi+\pi $. Before proceeding, we point out that the ``conformal invariance'' of the four-dimensional metric is equivalent to the shift symmetry $ \phi\rightarrow\phi+\pi $ of the scalar field $ \phi $. Let us now return to Einstein's second communication. The following argument used by Einstein would support our interpretation of eq. (\ref{invarianti}).

From Einstein's argument emerged clearly the compatibility of Kaluza's first postulate with conformal invariance, which he adopted as follows: `Let us suppose that only the ratios of the components of the metric tensor $ \gamma_{\bar{\mu}\bar{\nu}} $ have \textit{objective meaning} or --expressed differently-- [...] not the metric $ d\sigma^2 $, but only the totality of the ``null-cones'' $\left(  d\sigma^2 =0\right) $ is given.' [emphasis added] (\cite{E27b}; p. 27). Einstein tried to specify his idea by presenting it from a more geometrical point of view. In GR all the components of the metric tensor are needed in order to describe the four-dimensional reality. Einstein supposed that for the five-dimensional manifold it could be sufficient to determine the null-geodesics of the five-dimensional metric. Why then did Einstein choose this specific physical principle? Assuming that only the five-dimensional null-geodesics have physical meaning is equivalent to assume that a conformal transformation of the five-dimensional metric tensor does not have any influence on the four-dimensional physics, as stated above. As anticipated in the prologue, section \ref{prologue}, he was aware of the effects produced by the introduction of a physical extra-dimension, encoded by the effects of an extra scalar field to the four-dimensional dynamics. Einstein implicit statement on the invariance of the four-dimensional metric conformal transformations of the five-dimensional metric means that the equivalence class of all the five-dimensional conformally related metrics would produce the same four-dimensional physics. Therefore, Einstein extended the gauge principle presented in the first communication and  used this gauge freedom to justify the choice $ \gamma_{55} = 1 $. Indeed, the condition $ \gamma_{55} = 1 $ would correspond to a particular conformal frame obtained with the following transformation: $ {\gamma'}_{\bar{\mu}\bar{\nu}} = \Omega^2\gamma_{\bar{\mu}\bar{\nu}} $ with $ \Omega^2 = \left( \gamma_{55}\right)^{-1}  $. From this point of view, Einstein's choice can be interpreted as a sort of gauge fixing in a modern sense. Our interpretation of Einstein's approach to the construction of his unified action is supported also by the following fact. In 1927 Norbert Wiener and Jan Dirk Struik published an attempt to construct a four-dimensional unified theory for gravitational and electromagnetic interactions which would emerge from a Schr\"odinger equation \cite{WS-27} and \cite{WS-28}. The two authors started from a second order partial differential equation and defined a metric by imposing the conformal covariance of the equation itself\footnote{We shall not enter into technical details, because the work of Wiener and Struik will be investigated in a forthcoming paper.}. In modern language, they obtained a conformally invariant KG equation in a curved background\footnote{They published for the first time what we call nowadays the Penrose-Chernikov-Tagirov equation.}. Wiener and Struik considered also the possibility of introducing a fifth dimension by comparing their approach with Klein's first paper. In \cite{WS-27}, the authors emphasized the incompatibility between the two attempts, but in \cite{WS-28} they changed their point of view. Furthermore, they explicitly referred to Einstein's paper emphasizing the role played by the conformal transformations, which they called ``normalization theory'', in Einstein's attempt to justify the sharpened cylinder condition: `Although we [...] rejected the five-dimensional
theory [...], we reintroduce it here [...] \textit{we regard it [the normalization theory] as a necessary tool for the further investigation of the problems here considered, more especially if it is desired to reduce the number of the
fundamental constants of physics to a minimum. It has a particular bearing
on the discussion of the constancy of $ \gamma_{55} $ \cite{E27a} \cite{E27b}.}' [emphasis added] (\cite{WS-28}; p. 267). Therefore, also Wiener and Struik recognized an attempt of unifying Kaluza's theory with conformal transformations in Eintein's paper.  

As already stated in our Prologue, section \ref{prologue}, after having discussed the possibility of setting $ \gamma_{55}=\text{constant} $, Klein made a similar statement on the ratios of $ \gamma_{\bar{\mu}\bar{\nu}} $. Did Einstein simply replicate Klein's idea? 
In section \ref{merge}, we argue that Einstein had already considered the possibility of considering the postulate on light-cones as a viable principle for extending the realm of GR. Hence, we could say he was the father of this approach, Both Klein and Einstein introduced Weyl's idea to justify, from a physical point of view, the choice $ \gamma_{55} = constant $. In the following we shall see that Einstein would explicitly emphasize that this principle should also be used in order to construct the five-dimensional Lagrangian density, while Klein assumed without any further justification that the five-dimensional curvature scalar would be a viable Lagrangian density.

In the second paragraph of \cite{E27b}, Einstein showed that projecting the five-dimensional geodesics onto the hyperplane $ x^5 = \text{constant} $ yields the four-dimensional geodesic for charged particles in an electromagnetic and gravitational field, i.e. the Lorentz force law in four dimensions. Unlike Klein \cite{Klein1} and Fock \cite{Fock}, quoted at the end of this second communication thanks to Mandel's interest, Einstein did not considered null-geodesics only, even if he have previously stated that only the five-dimensional null-geodesics should have an objective meaning. Indeed, he obtained the Lorentz equation from the following Euler-Lagrange equations, namely:
\begin{equation}\label{geodesics-Einstein}
\delta\left( \int W_E d\tau \right) = 0\; ,
\end{equation}
where
\begin{equation}\label{WE}
W_E = \sqrt{ g_{\mu\nu}\frac{dx^\mu}{d\tau}\frac{dx^\nu}{d\tau}+\left(  \frac{dx^5}{d\tau}+A_\mu\frac{dx^\mu}{d\tau}\right)^2}\; ,
\end{equation}
while Klein introduced a different Lagrangian\footnote{Klein's Lagrangian can be be used for arbitrary geodesics, but in the case of massless particles $ \tau $ cannot be identified with the proper time \cite{MTW}. We remember that $ \kappa = 1 $ and that therefore $ W_E $ is dimensionless.}, namely $ \displaystyle L = \frac{1}{2}W_E^2 $. Einstein's postulate on the reality of null-cones is represented, as observed in \cite{Einstein-papers15bis}, by the $ \gamma_{55}=1 $ condition. Therefore, Einstein's approach was different from Klein's first attempt, but, as Einstein himself admitted, no new results can be obtained. It is worth remembering that Klein and Fock, followed an explicitly analogy with light in their first paper. But Klein emphasized the fact that Lorentz equation can be obtained also by projecting an arbitrary geodesic line in \cite{Klein2}, where the author presented also the quantization of the electric charge as a consequence of the compactification of the extra dimension. As Halpern emphasized, Einstein was familiar with Klein's work (\cite{Halpern}; p. 398)  and he was also aware of Klein's idea about a possible explanation for the quantization of the electric charge\footnote{In \cite{Einstein-papers15bis} an English translations of the original letters can be found} \cite{cite-Letter-14279}; \cite{Einstein-papers15bis}, an idea that would publish on \textit{Nature} \cite{Klein2}. But in (\cite{cite-Letter-14279}; \cite{Einstein-papers15bis}), Klein explicitly declared that he wanted to connect this last idea with Schr\"odinger's wave mechanics, which he criticized for the introduction of the $ n- $dimensional configuration space (\cite{cite-Letter-10145}; \cite{Einstein-papers15bis}). Furthermore we know that in January 11, 1927, before presenting his communications, Einstein wrote to Ehrenfest `My heart doesn't warm this Schr\"odingerism [...] \textit{I don't believe that kinematics must be abandoned}.'[emphasis added] (\cite{Einstein-papers15bis}; p. 447). Therefore, it is not surprising that Einstein did not read Klein's short communication to \textit{Nature} and that he did not yet explored the possibility of obtaining Lorentz equation from arbitrary geodesics. This fact forced Goenner to write, about this Klein's paper, that it `seems to have escaped Einstein' (\cite{Goenner}; p. 65). The emphasis in Einstein's answer to Ehrenfest enforces Halpern's idea that `Einstein had begun to disassociate himself
from the emerging Copenhagen probabilistic interpretation [and that] he wanted to
keep his distance from them' (\cite{Halpern}; p. 399): hence, he did neither quote Klein's first paper, as Halpern suggested, nor, we suggest, read Klein's short note to \textit{Nature}. Furthermore, the emphasis in Einstein's comment confirms, like suggested also by Jeroen van Dongen\footnote{Private communication.}, the important role played by the concept of particle in this period, which should not be abandoned in favour of the new mechanics. Indeed, soon before the communications we are discussing, Einstein and Grommer had published a work, i.e. \cite{EG-27}, where they analysed `the relationship between the field equations and the equation of motions' (\cite{Vizgin}; p. 221). As van Dongen emphasized, `It is natural to
wonder how Einstein would hope to undercut quantum theory with classical
Kaluza--Klein theory. [...] However, in none of the papers, or in his correspondence, is there
any explicit mention of how this should come about.' (\cite{VanDongen2}; p. 194). Furthermore, as Vizgin observed: `In his paper of 1927, Einstein ignored all aspects of the fifth dimension associated with quantum mechanics, probably because he hoped to obtain particles as singular solutions of unified field equations and the quantum features of their behaviour as properties of these solutions.' (\cite{Vizgin}; p. 232). We point out that Einstein had already identified the Weyl's conformal transformations as a `logical possibility' for extending GR in order to include microscopic phenomena\footnote{See section \ref{new-idea}.}. Einstein's statement on the status of particle solutions, his refuse of wave mechanics and his ideas on conformal transformations suggest the idea that he hoped to use a new physical principle, viz. Weyl's conformal invariance, in order to describe the path of microscopic particles in the context of Kaluza's theory\footnote{As it is well known, Einstein pointed out the inability of Kaluza's original theory of descibing electron's motion.}. 

Einstein knew that the introduction of the scalar field would produce a non-trivial modification of the Lorentz force law, but following our interpretation on the role of conformal transformations, he was convinced that his results should be invariant under conformal transformations of the five-dimensional metric. If Einstein would have considered the line element (\ref{general-metric}) with $ \alpha = \beta $, he would have noticed that $ \gamma_{55}=1 $ is only a sufficient condition. Indeed, using eq. (\ref{general-metric}) we can repeat Einstein's arguments as follows. By assuming $ \tau $ be an arbitrary parameter in five dimensions, the geodesic lines are characterized by the equation:
\begin{equation}\label{geodesics}
\delta\left( \int W d\tau \right) = 0\; ,
\end{equation}
where
\begin{equation}\label{W}
W = \sqrt{e^{2\beta\phi}\left[ g_{\mu\nu}\frac{dx^\mu}{d\tau}\frac{dx^\nu}{d\tau}+\left(  \frac{dx^5}{d\tau}+A_\mu\frac{dx^\mu}{d\tau}\right)^2\right]}\; .
\end{equation}
Following Einstein, we can choose $ \tau $ so that $ W = const. $ and the fifth component of the Euler-Lagrange equations following from eq. (\ref{geodesics}) reads:
\begin{equation}\label{eq-cons}
\frac{d}{d\tau}\left( \frac{\partial W}{\partial \dot{x}^5}\right) =
 \frac{d}{d\tau}\left( \frac{1}{2W}2e^{2\beta\phi}\left( \dot{x}^5+A_\mu\dot{x}^\mu\right) \right) = 0  \; ,
\end{equation} 
where we defined $ \displaystyle \dot{x}^{\bar{\mu}} = \frac{dx^{\bar{\mu}}}{d\tau} $. Then, eq. (\ref{eq-cons}) yields 
\begin{equation}
e^{2\beta\phi}\left( \dot{x}^5+A_\mu\dot{x}^\mu\right)  = A\; ,
\end{equation}
which coincides with Einstein's $ A $ (\cite{E27b}; p. 27) when $ \beta = 0 $, which is equivalent to $ \phi = 0 $, i.e. when $\gamma_{55} = 1 $, but it is not invariant under conformal transformations of the five-dimensional metric. At this stage, Einstein was able to deduce $  \displaystyle g_{\mu\nu}\frac{dx^\mu}{d\tau}\frac{dx^\nu}{d\tau} = constant $ and to set the constant equals to $ -1 $ in order to identify $ \tau $ with the four-dimensional proper time, but from eq. (\ref{W}) it follows that
\begin{equation}
g_{\mu\nu}\frac{dx^\mu}{d\tau}\frac{dx^\nu}{d\tau} = W^2e^{-2\beta\phi} - A^2e^{-4\beta\phi}\; ,
\end{equation}  
which is constant on the geodesic lines only if the scalar field $ \phi $ is constant on the geodesic lines, namely
\begin{equation}
0 = \frac{d}{d\tau}\left( e^{-2\beta\phi}\right) = e^{-2\beta\phi}\left( \dot{x}^\mu \partial_{\mu}\phi\right)  = e^{-2\beta\phi}\left( \dot{x}^\mu \nabla_{\mu}\phi \right)  \; ,
\end{equation} 
where we used the scalar character of $ \phi $ in the last equality. This additional condition would imply that $ \gamma_{55} $ would be constant on the geodesic lines only, not necessarily on the whole five-dimensional space. The constancy of $ \gamma_{55} $ is assured on the Killing trajectories, therefore a sufficient condition is to ask that the Killing trajectories be geodesic lines, which is equivalent to impose, as Einstein requested, the constancy of $ \gamma_{55} $ in the whole five-dimensional manifold, i.e the sharpened cylinder condition, as we already pointed out in section \ref{Ein-27a}. Finally, by noticing that Einstein's function $ W_E $ is related with $ W $ by $ \displaystyle W = e^{\beta\phi}W_E$, the other four equations following from eq. (\ref{geodesics}) read:
\begin{equation}\label{new-Lorentz}
\frac{d}{d\tau}\left( \frac{\partial W}{\partial \dot{x}^\mu}\right) - \frac{\partial W}{\partial x^\mu} = 
\frac{d}{d\tau}\left( e^{\beta\phi} \frac{\partial W_E}{\partial \dot{x}^\mu}\right) - \frac{\partial \left(  e^{\beta\phi}W_E\right) }{\partial x^\mu}=
e^{\beta\phi}\left[ \frac{d}{d\tau}\left(  \frac{\partial W_E}{\partial \dot{x}^\mu}\right) - \frac{\partial  W_E}{\partial x^\mu} - W_E\frac{\partial \left( \beta\phi\right) }{\partial x^\mu}\right]  =0\; .
\end{equation}
In order to obtain the Lorentz equation from eq. (\ref{new-Lorentz}), i.e. the first two terms of the squared bracket, we need an additional hypothesis: the quantities $ \partial_{\mu}\phi $ should be zero along five-dimensional the geodesic lines\footnote{In the case of null-geodesics, by using Klein's Lagrangian, a similar equation would hold, where the third term in the squared bracket of eq. (\ref{new-Lorentz}) would vanish, because of the constraint $ W_E = 0 $.}. Once again, the constancy of $ \gamma_{55} $ is a sufficient condition. This also shows that the Lorentz equation cannot be invariant under an arbitrary conformal transformation $ \phi\rightarrow\phi + \pi $.

In the third part of the communication, Einstein constructed his Lagrangian density for determining the field dynamics, which is equivalent to Klein's choice, i.e. the five-dimensional curvature scalar, up to a total derivative. Indeed, Einstein introduced the following Lagrangian density:
\begin{equation}\label{Ein-lagr-27}
\tilde{\mathcal{L}}=\sqrt{-\gamma}\gamma^{\bar{\mu}\bar{\nu}}\left[ \tilde{\Gamma}_{\bar{\mu}\bar{\nu}}^{\bar{\alpha}} \tilde{\Gamma}_{{\bar{\alpha}\bar{\beta}}}^{\bar{\beta}} -	\tilde{\Gamma}_{\bar{\alpha}\bar{\mu}}^{\bar{\beta}} \tilde{\Gamma}_{{\bar{\nu}\bar{\beta}}}^{\bar{\alpha}}\right]  \; ,
\end{equation}
where the five-dimensional Christoffel symbols are defined, as usual in GR, by using the metric tensor $\gamma_{\bar{\mu}\bar{\nu}}  $ and its first derivative and he specified that eq. (\ref{Ein-lagr-27}) `is expressed using $ g_{\mu\nu} $ and $ A_\mu $' (\cite{E27b}; p. 28). This statement follows after that Einstein fixed $ \gamma_{55}=1 $ again, hence, $ g_{\mu\nu} $ and $ A_\mu $ should be identified with equations (\ref{var1}) and (\ref{var2}). But this statement can be interpreted also as an emphasis to the fact that only $ g_{\mu\nu} $ and $ A_\mu $ are physical variables, despite of the five-dimensional approach. Unlike Klein, Einstein did not try to connect the fifth dimension with some physical feature of the model.  Our interpretation of the role of conformal transformations in Einstein's approach is coherent with the idea that Einstein interpreted the fifth dimension, in 1927, only as a mathematical tool, as he had already suggested\footnote{See section \ref{5-dim}.}. Once again, we emphasize that Einstein was aware of the phenomenological problems created by the introduction of an extra scalar field associated to a physical meaning of the fifth dimension. The fact that a conformal transformation should not have any effect on the four-dimensional physics means that in 1927, the fifth dimension should have no physical significance for Einstein.

At the end of the second communication, Einstein showed that equation (\ref{Ein-lagr-27}) reduces to the sum of the usual Einstein-Hilbert and Maxwell Lagrangian densities, namely
\begin{equation}\label{lagr-dens}
\tilde{\mathcal{L}}=\sqrt{-g}\left[ g^{\mu\nu}\left(  \Gamma_{\mu\nu}^{\alpha} \Gamma_{\alpha\beta}^{\beta} -	\Gamma_{\alpha\mu}^{\beta} \Gamma_{\nu\beta}^{\alpha}\right) -\frac{1}{4}   F_{\mu\nu}F^{\mu\nu}  \right] \; .
\end{equation}
Once again, Einstein obtained the result presented by Klein, but considering the difference between Einstein's and Klein's choice for the four-dimensional metric, which is similar to Klein's work, Einstein's result suggests the following observations. By identifying the four-dimensional metric of eq. (\ref{lagr-dens}) with $ g^E_{\mu\nu} $ and setting $ \displaystyle A_\mu=\frac{\gamma_{5\mu}}{\gamma_{55}} $, i.e. the first two expressions of eq. (\ref{invarianti}), the Lagrangian density (\ref{lagr-dens}) is invariant under conformal transformations of the five-dimensional metric\footnote{The inverse components of $ g_{\mu\nu} $ read $ g^{\mu\nu}= \gamma_{55}\gamma^{\mu\nu} $, which are also conformally invariant quantities as well as the determinant of the four-dimensional metric} and the fifth dimension looses its physical meaning, but it no longer corresponds to the five-dimensional Einstein-Hilbert action. In order to understand how this new conformally invariant action would look like, we could perform a conformal transformation of the five-dimensional metric. But we are interested in showing how the action can be written from a five-dimensional point of view, therefore we start considering the dimensional reduction obtained with arbitrary values of $ \alpha $ and $ \beta $ (introduced before in our description of Klein and Einstein results). After straightforward calculations, the Einstein-Hilbert Lagrangian reads:
\begin{equation}\label{red1}
\tilde{R}\sqrt{-\gamma} = \sqrt{-g}e^{\left( \beta+2\alpha\right) \phi}\left[ R -2\left( 3\alpha +\beta\right)\square\phi - \left( 6\alpha^2+4\alpha\beta +2\beta^2 \right) g^{\mu\nu}\partial_{\mu}\phi\partial_{\nu}  \phi -\frac{1}{4} e^{2\left( \beta-\alpha\right)\phi }F^2  \right] \; ,
\end{equation} 
where $\displaystyle \square\phi = \frac{1}{\sqrt{-g}}\partial_{\mu}\left( \sqrt{-g}\partial_{\mu}\phi\right)  $ and $ \displaystyle F^2 = F_{\mu\nu}F^{\mu\nu}$.
First, we notice that Einstein's choice $ \alpha = \beta $ would be an unusual choice also in modern Kaluza-Klein theories \cite{Raife2000}, because only if $ \beta = -2\alpha $ the gravitational part of eq. (\ref{red1}) would produce the four-dimensional Einstein-Hilbert action. Indeed, after setting  $ \alpha = \beta $, eq. (\ref{red1}) reads:
\begin{equation}\label{red2}
\tilde{R}\sqrt{-\gamma} = \sqrt{-g_E}e^{3\beta\phi}\left[ R_E -8\beta\square_E\phi - 12\beta^2 g_E^{\mu\nu}\partial_{\mu}\phi\partial_{\nu}  \phi -\frac{1}{4}F^2_E  \right] \; ,
\end{equation} 
where the index $ E $ simply indicates that the four-dimensional metric and the electromagnetic potentials are defined like in eq. (\ref{invarianti}) and we remember that $ \gamma_{55} = e^{2\beta\phi} $. By inverting eq. (\ref{red2}) we get the desired action, namely:
\begin{eqnarray}\label{action-inv}
\sqrt{-g_E}\left[ R_E  -\frac{1}{4}F^2_E  \right] &=&e^{-3\beta\phi}\tilde{R}\sqrt{-\gamma}
 + 12\beta^2 g_E^{\mu\nu}\partial_{\mu}\phi\partial_{\nu}  \phi +8\beta\square_E\phi
\nonumber\\
&=& e^{-3\beta\phi}\sqrt{-\gamma}\left[ \tilde{R} + 12\beta^2 \gamma^{\mu\nu}\partial_{\mu}\phi\partial_{\nu}  \phi \right] + t.d. \; ,
\end{eqnarray}
where we have discarded a total derivative and we have used the following identities: $ \displaystyle \sqrt{-g_E} = e^{-5\beta\phi}\sqrt{-\gamma} $ and $ \displaystyle g_E^{\mu\nu}=e^{2\beta\phi}\gamma^{\mu\nu} $. The action obtained is invariant under conformal transformations of the five-dimensional metric by definition, but its invariance can be checked directly using the following identities\footnote{After an integration by parts, the transformed action is equivalent to the untransformed one up to a total derivative.}: $ \phi'=\phi + \pi $; $ \gamma_{\bar{\mu}\bar{\nu}}' = e^{2\beta\pi}\gamma_{\bar{\mu}\bar{\nu}} $; $ \tilde{R}' = e^{2\beta\pi}\left[ \tilde{R} -8\beta\square\pi - 12\beta^2\gamma^{\mu\nu}\partial_{\mu}\pi\partial_{\nu}  \pi \right]  $. From a modern point of view, this action would resemble a non-linear scalar field action non-minimally coupled with gravity, but the scalar field $ \phi $ would not transform as a canonical scalar field under conformal transformations. If we define $\displaystyle \psi = \sqrt{\xi}e^{-\frac{3}{2}\beta\phi}$, where $ \xi\in \mathbb{R} $ is an appropriate constant, the new scalar field $ \psi $ would transform as a scalar field in five dimensions\footnote{When $ \gamma_{\bar{\mu}\bar{\nu}}\rightarrow \Omega^2\gamma_{\bar{\mu}\bar{\nu}} $, in $ n- $dimensions a scalar field would transform as follows: $ \psi\rightarrow\Omega^{\frac{2-n}{2}}\psi $. For $ n = 5 $, the definition $\displaystyle \psi = \sqrt{\xi} e^{-\frac{3}{2}\beta\phi} $ yields  $ \psi\rightarrow\Omega^{-\frac{3}{2}}\psi $.} and the action reads:
\begin{equation}\label{action}
\sqrt{-\gamma}\left[ \frac{1}{\xi}\psi^2\tilde{R} + \frac{1}{2}\gamma^{\bar{\mu}\bar{\nu}}\partial_{\bar{\mu}}\psi\partial_{\bar{\nu}}{\psi} \right]\; .
\end{equation}
Equation (\ref{action}) would resemble a phantom scalar field, because of the wrong plus sign in front of the kinetic part for the scalar field, non-minimally coupled with gravity in five dimensions, but it should be intended as a constrained Lagrangian, because $ \psi $ and $ \gamma_{55} $ are not independent variables. Now, we return to Einstein's communications.

At the end of his work, as also Vizgin observed (\cite{Vizgin}; p. 232), Einstein discussed the possibility of setting $ \gamma_{55} = -1 $, i.e. the possibility of choosing a time-like, instead of a space-like, extra dimension, a choice that Kaluza also tried to discuss, as we will point out in the following section. With a time-like extra-dimension, a wrong sign between the gravitational and the electromagnetic Lagrangian densities would appear in eq. (\ref{lagr-dens}). Hence, Einstein concluded that this fact forced the extra-dimension to be space-like, instead of time-like. We emphasize that this fact emerged explicitly for the first time in the history of the Unified Field Theories. Indeed, we are aware of the fact that a similar discussion would take place in the same year between Klein and Louis de Broglie, after de Broglie tried to introduce a time-like extra-dimension, and that Klein would publish a similar discussion at the end of the year (\cite{Klein5}; p. 206); (\cite{Peruzzi-Rocci-Rosenfeld}; p. 199-200).

\section{Einstein-Kaluza correspondence revisited}\label{Kaluza}
The modern multidimensional theories are often called \textit{Kaluza-Klein theories}. Indeed, before Klein, Theodore Kaluza introduced also a five-dimensional space-time \cite{Kaluza} \cite{Kaluza-trad}, in order to unify GR and EM\footnote{Gunnar Nordstr\"om also considered a five-dimensional space-time before Kaluza \cite{Nord}, but the Norwegian mathematician described the gravitational interaction using a scalar field instead of a tensor field.}. Kaluza's theory has been largely analysed from the historical point of view, see e.g. \cite{Vizgin}, but a brief review is needed, in order to set the background and the various stages of the Kaluza-Einstein discussion. Kaluza sent Einstein a draft of his paper in 1919, but Einstein communicated Kaluza's work only 2 years later. Why did Einstein wait for so long time? It is well known that Einstein pointed out some unwanted features of Kaluza's model. The aim of this section is to analyse the role played by the cylinder condition both during the beginning of Einstein-Kaluza correspondence and in his following works.

Kaluza introduced the fifth dimension interpreting the electromagnetic tensor $ F_{\alpha\beta} $ as a sort of truncated Christoffel symbol. Hence, using five-dimensional Christoffel symbols $  \tilde{\Gamma}^{\bar\nu}_{\bar\mu\bar\sigma} $, Kaluza was able to introduce both gravitational and electromagnetic potentials. In order to take into account the non observability of the fifth dimension, Kaluza introduced the cylinder condition, stating that: `one has to take into account the fact that we are only aware of the space-time variation of state-quantities, by making the derivatives with respect to the new parameter vanish or by considering them to be small as they are of higher order' (\cite{Kaluza}; p. 967), which is equivalent to the cylinder condition, namely $\partial_5\gamma_{\bar{\mu}\bar{\nu}} = 0 $.   

Unlike Klein, Kaluza treated $ \gamma_{55} $ as variable, even if he considered the linearised five-dimensional Einstein equations only, i.e. in the weak-field limit. Kaluza called this hypothesis \textit{approximation I}. In Kaluza's paper there is no explicit discussion of the full field equations. The author considered the fifth dimension as a physical ingredient of his theory, because he discussed the consequences implied by the introduction of a non-constant $ \gamma_{55} $ for our four-dimensional world. Kaluza was concerned with the particle's geodesic motion in five dimensions, which should be connected, in Kaluza's theory, with the motion of charged particles in four dimensions in the presence of gravitational and electromagnetic interactions. It is worth noting that in the concluding paragraph of his paper, Kaluza maintained a neutral position with respect to the physical meaning of the new ``world parameter'', i.e. the fifth dimension. As Kaluza said, he encountered `physical as well epistemological difficulties' (\cite{Kaluza}; p. 967) in giving a physical meaning to the new formalism. Kaluza's difficulties are described in the following.   

Kaluza investigated the five-dimensional geodesic motion in the small-velocities limit and called it \textit{approximation II}. Thanks to this approximation, an extra term coming from the non-constancy of  $ \gamma_{55} $ disappeared from the geodesic equation. Hence, the five-dimensional geodesics corresponded to the Lorentz equation for charged particles in the presence of gravitational and electromagnetic interaction. The small-velocities approximation implied also that the charged matter should have `a very tiny specific charge $ \rho_0 / \mu_0 $' (\cite{Kaluza}; p. 969), where $ \rho_0 $ and $ \mu_0 $ are the electric charge and the particle's mass and their ratio is what Kaluza called the specific charge. Exploring the possibility of applying his model to microscopic phenomena, Kaluza emphasized that Einstein has pointed out an inconsistency of his model in this context (\cite{Kaluza}; p. 971). Indeed, approximation II cannot be satisfied by electrons\footnote{Using the modern values of electron's charge and mass, the order of the ratio is (IS) $\sim 10^{11} \frac{C}{kg} $.}.

Kaluza's comment is important for two reasons. First, it emphasizes that the fifth dimension was connected with measurable quantities, even if it should not be observable by itself. Kaluza was afraid that the new quantum theory, concerning the microscopic phenomena, should threaten for the validity of his model. Second, it points out how Einstein was concerned with the physical meaning of the fifth dimension and with the physical consequences of Kaluza's approach. 

Was Einstein aware of the consequences of setting  $ \gamma_{55} =\text{constant}$ in 1921, for instance when he communicated Kaluza's paper to the \textit{Preussische Akademie}? In order to answer this question, we reconsidered Einstein-Kaluza correspondence between 1919, when Einstein received Kaluza's manuscript for the first time, and 1921, when Einstein convinced himself to communicate Kaluza's results.

\subsection{Kaluza-Einstein correspondence}\label{KE-correspondence}
Einstein first reaction to Kaluza's introduction of the extra-dimension was enthusiastic in April 1919. Indeed, Einstein wrote: `your idea [Kaluza's] has great appeal for me. It seems to me to have decidedly more promise from the physical point of view than the mathematically probing exploration by Weyl' (\cite{Einstein-papers9bis}; p. 21). In making his assertion, Einstein contrasted Kaluza's theory with Hermann Weyl's approach\footnote{We will investigate Weyl's role in section \ref{Weyl}.}. Einstein defined Weyl's theory like a `\textit{mathematically} probing exploration' for various reasons. For our purpose it is worth mentioning that Einstein pointed out that Weyl's theory was in contrast with the empirical facts. Furthermore, Weyl considered the potentials $ A_\mu $ as a fundamental objects, instead of electric and magnetic fields, but in Einstein's opinion they were physically meaningless quantities (\cite{Einstein-papers12bis}; p. 52) . On the contrary, Kaluza's idea is based on the electromagnetic tensor. 

In this first letter and in the following correspondence, it emerges how Einstein analysed the weaknesses of Kaluza's model. From our point of view, the most important comment is the following: `It now all depends on whether your idea will withstand on \textit{physical} scrutiny.' (\cite{Einstein-papers9bis}; p. 21). This comment emphasizes that Einstein was concerned with the measurable consequences of the introduction of the fifth dimension, as we will see in the following.

In his second answer to Kaluza, Einstein underlined again the importance of empirical proofs in considering the projection of five-dimensional geodesics onto four-dimensional slices $ x^5 = \text{constant} $. `I myself definitely would not publish the idea -- if it had occurred to me -- before having done this test, which seems to be a secure and relatively simply criterion.' (\cite{Einstein-papers9bis}; p. 26). In our opinion, this is the reason why Kaluza performed his approximations: he wanted to compare his mathematical approach with the empirical reality. In his third letter, Einstein convinced himself, with the help of Kaluza, that `from the point of view of realistic experiments, your [Kaluza's] theory has nothing to fear.' (\cite{Einstein-papers9bis}; p. 32).

Third Einstein's letter shall play an important role in the following section of our paper. Indeed, Einstein pointed out two important arguments against Kaluza's approach, that he would continue to grapple with these problems in his future developments of Kaluza's theory. First he noted the incompatibility between the request of the general covariance in five-dimensions and the cylinder condition. As we said, also in the published version of Kaluza's paper, the author made no comments about this fact. Second, Einstein judged as \textit{very unsatisfactory} the cylinder condition: `One requires: 1) General covariance in $ R_5 $. 2) In combination with this, the relation $\displaystyle \frac{\partial}{\partial x^5} =0 $ \textit{not} be \textit{covariant} in $ R_5 $. Obviously, this is very unsatisfactory' (\cite{Einstein-papers9bis}; p. 32). This comment answers one of our questions: Einstein was struggled from the very beginning by the fact that the cylinder condition was not formulated in a covariant form. 

In his fourth answer, Einstein pointed out finally the inconsistency of Kaluza's model, when applied to microscopic phenomena, i.e. the motion of an electron. `[...] upon more careful reflections about the consequences of your interpretation, I did hit upon another difficulty, which I have not been able to resolve until now.' (\cite{Einstein-papers9bis}; p. 36). In order to understand the limits of applicability of Kaluza's model, Einstein considered the field equations with a non-constant $ \gamma_{55} $. Indeed, Einstein claimed that he had calculated the full field equations in the first approximation, but he reported the $ 55- $component only. He used this equation for estimating the order of magnitude of $ \gamma_{55} $, and used it to estimate the magnitude of the fifth component of the particle's velocity in the case of an electron. Indeed, the two quantities appear in the fifth component of the geodesic equation. The huge order of magnitude he obtained was incompatible with the small-velocities approximation proposed by Kaluza for eliminating the physical effects of the fifth dimension.

Einstein's calculation assumes a new and important role for our purpose. Let us take a closer look to Einstein's letter. He considered a weak field limit like Kaluza (approximation I). Hence, in order to calculate the field equations, Einstein set, for the determinant of the metric, $ \gamma = 1 $ and used the following Lagrangian density
\begin{equation}\label{Ein-lagr}
\mathcal{L}=	\gamma^{\bar{\mu}\bar{\nu}}\tilde{\Gamma}_{\bar{\alpha}\bar{\mu}}^{\bar{\beta}} \tilde{\Gamma}_{{\bar{\nu}\bar{\beta}}}^{\bar{\alpha}} \; .
\end{equation}
The Lagrangian density (\ref{Ein-lagr}) is equivalent, up to a total derivative and in the weak-field limit, to the five-dimensional curvature scalar $ \tilde{R} $. Following Kaluza's Ansatz and using (\ref{Ein-lagr}), the $ 55- $component of Einstein equations reads (\cite{Einstein-papers9bis}; p. 36):
\begin{equation}\label{55-eq-E}
	\frac{1}{2}\square \gamma_{55} = \frac{\kappa^2}{4} {\hat{F}}_{\alpha\beta}{\hat{F}}^{\alpha\beta} \; ,
\end{equation}
where, Einstein adopted Kaluza's definition for the electromagnetic potentials $ \hat{A}_\mu = \gamma_{5\mu} $, which are related with Klein's definition of $ A_\mu $ as follows: $\displaystyle  A_{\mu}  = \frac{\hat{A}_\mu}{\gamma_{55}} $ (cf. with definition \ref{def-A}). By starting with r.h.s of equation (\ref{55-eq}), inserting $\displaystyle  A_{\mu}  = \frac{\hat{A}_\mu}{\gamma_{55}} $ in the definition of $ F_{\alpha\beta} $ and neglecting the non-linear terms, equation (\ref{55-eq}) reduces to (\ref{55-eq-E}). From eq. (\ref{55-eq-E}) Einstein would have been able to infer that $ \gamma_{55} =\text{constant} $ implied $ {\hat{F}}_{\alpha\beta}{\hat{F}}^{\alpha\beta}=0 $. Indeed, in 1919, he did not suggest this short cut to eliminate the term proportional to $ \partial_{\mu}\gamma_{55} $ from the geodesic equation, which describes the effect of $\gamma_{55} $ on four-dimensional particle's dynamic. As we said above, Kaluza eliminated this extra term using a different trick, at the price to accept the inapplicability of his model to the motion of electrons. In his subsequent answer to Kaluza (\cite{Einstein-papers9bis}; pp. 41-42), dated May 1919, Einstein emphasized again the importance of his objections, underlining that it would prevent him communicating Kaluza's paper to the Preussische Akademie. Notwithstanding this fact, Einstein considered Kaluza's model of mathematical interesting. He suggested that Kaluza submit the paper to a journal of mathematics (\textit{Matematische Zeitschrift}). Therefore, as it is well known, the physical meaning of the fifth dimension encoded by the unwanted features introduced by the scalar field, played an important role, for Einstein, in considering Kaluza's theory as physically unacceptable. Kaluza's draft of the paper remained unpublished for the following two years. 

In October 1921 Einstein decided to give a second chance to Kaluza's theory. Why did Einstein write again to Kaluza? To our knowledge, the question seems unanswered. It is worth noting that the idea of five dimensions was nominated by Jakob Grommer, Einstein's old collaborator, even if in a different contest. Indeed, at the time, Grommer was working on a new book on the theory of relativity. One of the aims of his work was `to make a few additions to [...] Weyl's invariants' (\cite{Einstein-papers12bis}; p. 145), but he posed many questions to Einstein, when he wrote to him on August 21. One of these questions regarded the maximal dimensionality of the manifold where our space-time can be embedded. In this context, Grommer emphasized that the dimension $ n $ should be greater than five, which did not work for the purpose he had in mind  (\cite{Einstein-papers12bis}; p. 146). We do not know Einstein's answers, but as a matter of fact, in his following letter to Einstein dated October 25, Grommer and Einstein had started a new collaboration by discussing on Kaluza's theory\footnote{Einstein and Grommer work would be published only at the beginning of 1923.} (\cite{Einstein-papers12}; p. 333). Nine days before, on October 14, Einstein had sent his letter to Kaluza. Now let us follow again Kaluza-Einstein correspondence. 

From the letter sent by Kaluza in response to Einstein's offer for a second chance, we know that during these years Kaluza tried to fix the problems that Einstein pointed out (\cite{Einstein-papers12bis}; p. 178). Even if he did not fix the problem of the magnitude of $ \gamma_{55} $, Kaluza declared `it does not appear to be quite insurmountably imposing as then'(\cite{Einstein-papers12bis}; p. 178). In a subsequent letter, Kaluza admitted he was not able to eliminate the discrepancy and proposed to Einstein the solution which appeared in the published paper. Kaluza also pointed out different proposals which occurred in his mind. Finally, Einstein communicated Kaluza's paper and it was published. In his last letter to Einstein before the publication of his paper, Kaluza mentioned that he tried to solve the problems created by the scalar field with different approaches. One was the introduction a \textit{time-like} extra-dimension (\cite{Einstein-papers12bis}; p. 191). At present we have no additional informations in order to investigate Kaluza's ideas, but from our point of view, the fact that Kaluza mentioned this possibility is implicitly connected with Einstein's discussion on the character of the fifth dimension.

\subsection{On the ontological status of the fifth dimension}\label{5-dim}
The unsolved points of Kaluza's paper were the applicability of the theory to the microscopic phenomena and the physical meaning of the fifth dimension connected with the introduction of the cylinder condition. One year after the publication of Kaluza's paper, in January 1922, Einstein and Grommer pointed out this two facts. They sent a communication to \textit{Scripta Universitatis atque Bibliothecae Hierosolymitanarum. Mathematica et Physica} \cite{EG-23} \cite{Einstein-papers13bis}, whose main purpose was to point out the non-existence of every-regular centrally symmetric field according to the theory of Kaluza\footnote{In the following we will refer to the English translation in \cite{Einstein-papers13bis}.}. What Einstein and Grommer meant was underlined by the authors in the final assertion of the paper: `Thus it is proven that Kaluza's theory possesses no centrally symmetric solutions dependent on the $ \gamma_{\bar{\mu}\bar{\nu}} $'s alone that could be interpreted as a (singularity free) electron.' (\cite{Einstein-papers13bis}; p. 33). With this comment, Einstein and Grommer emphasized the first point.

The second point emerged by reviewing Kaluza's theory. Indeed, the authors discussed the advantages of Kaluza's theory, briefly contrasting it with Weyl's approach, and pointed out its weakness. The advantages are summed up by the fact that Kaluza's theory offered a more natural way of unifying GR with Maxwell's theory from Einstein's point of view, because the two theories emerge from a unique five-dimensional Lagrangian density. The weaknesses are based on his critics about the non-covariant form of the cylinder condition, emerged in his correspondence with Kaluza, as the authors describe in the following. `In the general theory of relativity [...], the $ d\sigma^2 = \gamma_{\bar{\mu}\bar{\nu}}dx^{\bar{\mu}}dx^{\bar{\nu}} $ means a directly measurable magnitude for a local inertial system using measured rods and clocks, whereas the $ d\sigma^2 $ of the five-dimensional manifold in Kaluza's extension initially stands for a pure abstraction that \textit{seems not to deserve direct metrical significance}.' [emphasis added] (\cite{Einstein-papers13bis}; p. 31). Einstein and Grommer referred to the fact that the physical effects of the fifth dimension must be eliminated in order to obtain the usual four-dimensional particle's dynamic. The emphasis shows how Einstein started to loose his faith on the physical reality of the five-dimensional space-time and we can infer that he was wondering also what should have physical meaning in five dimensions. In addition, as they had underlined few lines before, the $ \gamma_{55} $ function still awaited interpretation. It is worth mentioning that in Einstein and Grommer's analysis $ \gamma_{55} $ played the role of a variable. The authors continued: `Therefore, from the physical point of view, the requirement of general covariance of all equations in the five-dimensional continuum appears completely unfounded.' (\cite{Einstein-papers13bis}; p. 31). As we underlined, Kaluza did not discuss the consequence that had the cylinder condition on the group of the coordinates transformations. Einstein and Grommer concluded: `Moreover, it is a questionable asymmetry that the requirement of the cylinder property distinguish one dimension above the others and yet with reference to the structure of the equations all five dimensions should be equivalent.' (\cite{Einstein-papers13bis}; p. 31). This last comment emphasizes once again the fact that the asymmetry implied by the the cylinder condition would enforce the interpretation of the fifth dimension as non-physical. Einstein would not change his mind until 1938, when he would explicitly pointed out that Kaluza's fifth dimension would have to be considered as physical \cite{Halpern} \cite{Witten}.

\subsection{The importance of being ``covariant''}
In the following years Einstein tried to develop different approaches in order to unify the gravitational and the electromagnetic interactions. His new starting point was to consider a generalized affine connection which would contain the Christoffel symbols as a particular case, an idea initiated by Weyl and Eddington, as it is widely explained in \cite{Goenner}. These attempts failed for various reasons. Like Vizgin writes: `he was clearly very disappointed in the efforts 
so far made to create unified geometrized field theories.' (\cite{Vizgin}; p. 219) Hence, `having lost faith in the affine and affine-metric theories related to the theories of Weyl and Eddington, Einstein returned to Kaluza's five-dimensional scheme.' (\cite{Vizgin}; p. 220). In addition, Vizgin observes that `it was an entirely natural step for Einstein to turn to the five-dimensional approach. Of course, this may have been due to the revival of five-dimensional field theory in connection with quantum mechanics.' (\cite{Vizgin}; p. 221). Goenner reported that `Einstein became interested in Kaluza's theory again due to O. Klein's paper' (\cite{Goenner}; p. 64), i.e. \cite{Klein1}, and that, as a consequence, Einstein wrote to his friend and colleague Paul Ehrenfest. It was Ehrenfest himself who drew Einstein's attention for the first time on Klein's work, by inviting Einstein to a discussion with Klein himself, but Einstein declined for personal reasons. Indeed, Klein visited Ehrenfest in Leiden soon after the cited letter, in order to give a talk on his ideas and Ehrenfest worried about the absence of Einstein, as he wrote in August 26 \cite{cite-Letter-10143} \cite{Einstein-papers15bis}. After Ehrenfest, Grommer attracted again Einstein's attention on first Klein's paper and Einstein asked to Ehrenfest a copy of Klein's work \cite{cite-Letter-10141} \cite{Einstein-papers15bis}. As a consequence, Einstein corresponded with Klein who outlined the future developments he tried to achieve with his five-dimensional approach, i.e. to connect the periodicity of the fifth coordinate with the quantization of the electric charge \cite{cite-Letter-14279} \cite{Einstein-papers15bis}. It is worth noting that Klein's letter was sent in August 29, while  Klein sent his note to \textit{Nature} five days later, September 3. Vizgin writes `From the middle of 1926, there had appeared in 
the\textit{ Zeitschrifit f\"ur Physik} alone not less than ten papers on the application of 
the five-dimensional approach to quantum mechanics. It appears that  
Einstein did not pay attention to these papers, since otherwise he would not have 
failed to mention the new aspect of the five-dimensional approach discovered 
by O. Klein and Fock' (\cite{Vizgin}; p. 221). Vizgin refers to the note added in proof \cite{E27b}, that we commented in section \ref{Ein-27}. The contents of the letters confirms that Einstein was aware of both the five-dimensional generalization of wave mechanics and the work of Klein. As we argued, he did not mention Klein's work, because he used a different approach, despite the fact that the results are the same.

After having received Klein's letter and having read Klein's paper on five-dimensional world, Einstein realized that not much had changed about the state of the cylinder condition since he stopped working on it. Corresponding with Ehrenfest, Einstein emphasized that `Klein's paper is good, but that Kaluza's theory is too unnatural' (\cite{Pais}; p. 350). A closer inspection of the postcard \cite{cite-Letter-10147} \cite{Einstein-papers15bis} showed that Einstein pointed out once again his disappointment regarding the contrast between the request of general covariance and the cylinder condition, defining it as \textit{unlikely}. This comment is dated September 3. Hence, we can infer that at that time he started again to investigate how to write the cylinder condition in a covariant form. At the time, Einstein was working with Grommer on the relationship between the field equations and the equation of motions \cite{EG-27}, which will be published at the beginning of 1927 in the same volume where the two communications on Kaluza appeared. This means Einstein realized that the cylinder condition can be formulated in a covariant form in this period. This fact and the idea of merging Kaluza's approach with Weyl's idea, an aspect that we shall consider in the next section, convinced Einstein to write the following statement to Lorentz on February 16th, four days before the Prussian Academy would receive the second communication \cite{E27b}: `It appears that the union of gravitation and Maxwell's theory is achieved in a completely satisfactory way by the five-dimensional theory (Kaluza-Klein-Fock).' (\cite{Goenner}; 65). This positive comment was implicitly motivated by the fact that he found a covariant formulation of the cylinder condition. Furthermore, from our point of view, there is a connection between the new formulation of the cylinder condition and Einstein's decision to publish his communications, as we argue in the following. 

In the process of creating GR, Einstein achieved the covariant form of the gravitational field equations with hundreds of stop and go. He had also formulated an argument \textit{ad hoc}, the already mentioned ``hole argument'' before 1915, when he was convinced that such form cannot exist. As Norton emphasized, from Einstein's point of view the principle of general covariance was `a principle with significant physical content, and [...] that content is the character of a generalized relativity principle.' (\cite{Norton-1}; p. 283). Indeed, in October 1916, Einstein wrote that the principle of equivalence is always satisfied if equations are covariant (\cite{Einstein-papers6bis}; p. 239). Once achieved, the principle of general covariance was an essential character of physical laws. Indeed, in his founding paper of GR, Einstein remarked that `the laws of Nature have to be expressed by generally covariant equations' (\cite{Einstein-papers6bis}; p. 153). Furthermore, Norton asserted: `Einstein also predicated the covariance property not directly to the model set but to the equations that define the model set [...] if the equations defining a model set are covariant under a group \textit{G} then the model set must also be covariant under that group and vice versa.' (\cite{Norton-1}; p. 292). Hence, the general covariance was an essential character both for all of the equations defining GR and for searching for new theories. As a consequence, the new formulation of the cylinder condition he published in 1927 can be interpreted as a little step forward to the formulation of a possible unified theory and could be advocated as one of the reasons that forced Einstein to publish his communications.

Before proceeding it is worth mentioning that in \cite{Halpern} the author propose another additional explanation. Halpern stated that `By that time, Einstein had begun to disassociate himself
from the emerging Copenhagen probabilistic interpretation of quantum mechanics,
as would become clear that October at the fifth Solvay Congress in Brussels. Perhaps,
therefore, although Einstein had been impressed by Klein's ideas, he wanted to
keep his distance from them, as they were linked to the emerging Copenhagen interpretation. [...] Kaluza--Klein theory offered a deterministic alternative to
probabilistic quantum mechanics' (\cite{Halpern}; p. 399). As we shall argue in section \ref{Weyl}, the role of conformal transformations we analyzed in section \ref{Ein-27} was connected with Einstein's project to find a deterministic alternative for describing electron's dynamics.

\section{Einstein, Weyl and the scale invariance}\label{Weyl}
Having investigated the role played by Kaluza's theory, the following questions are yet unanswered. Did Einstein want to unify electromagnetic and gravitational forces only in his brief communications or did he want to incorporate the microscopic phenomena also?  Why did Einstein choose the scale invariance as a physical principle for generalizing GR? In order to answer these questions we shall reconsider Einstein-Weyl correspondence between 1918 and 1921.

\subsection{Weyl's theory}
A brief review of Weyl's theory in a historical context can be found in \cite{Raife2000}. Here we summarize the most important aspects. Weyl's starting point was purely mathematical. He introduced an affine connection, which generalized the Levi-Civita connection. The geometrical idea is related to the concept of parallel transportation. In GR, the parallel transportation of a vector from a point to another could result in a rotation of the vector, but its modulus remains unchanged. Using Weyl's connection, the magnitude of the vector's modulus change. One advantage of Weyl's theory is that the new connection's symbols contain both the gravitational potential and a four-vector whose components can be interpreted as the electromagnetic potentials. As a consequence of these assumptions, using modern language, the four-dimensional space-time manifold of Weyl's theory is equipped with a conformal structure, i.e. with a set of conformally equivalent Lorentz metrics and not with a definite metric as in GR, a sort of new internal symmetry group. The idea of gauge-invariance emerged in those years, but the meaning attributed by the author was different from our modern concept (see \cite{DawningGauge} for further details). Before giving up his ideas, Weyl proposed many Lagrangian densities in order to achieve a unified theory of gravity and electromagnetism. All of them were conformally invariant objects: in this sense Weyl used an approach analogous to the modern idea of gauge-invariance.

The essential weakness of Weyl's theory was pointed out by Einstein very early. The main consequence of the introduction of the conformal symmetry is that the behaviour of clocks would depend on their history, a fact that does not have any empirical justification. From an epistemological point of view, what disturbed Einstein was that the line element $ ds $ lost the physical meaning which he had in GR: `I reported to you more exactly the objection bothering me with regard to your new theory. (Objective meaning for $ ds $, not just for the ratios of different $ ds $'s originating from one point.)' (\cite{Einstein-papers8bis}; p. 532). Indeed, as Weyls himself underlined, in his conformal invariant theory `only the ratios of the components of the metric tensor [...] have  a direct physical meaning' (\cite{DawningGauge}, p. 26).

One of the reasons that motivated Weyl was the idea to extend the realm of GR to the microscopic scales. Indeed, Weyl's purpose was to construct what he called `a true infinitesimal geometry' (\cite{DawningGauge}; p. 25). Einstein had tried by the beginning of 1919 to use GR for constructing a theory describing electrons \cite{Ein19} in a paper entitled \textit{Do Gravitational Fields Play an Essential Role in the Structure of the Elementary Particles of Matter?}  (\cite{Einstein-papers7bis}; English translation, p. 80). At the end of 1920 he wrote a contribution for \textit{Nature} on the development of the theory of relativity where he explicitly posed the following question. `Do gravitational fields play a part in the constitution of matter, and is the continuum within the atomic nucleus to be regarded as appreciably non-Euclidean?' ( \cite{Ein21}; p. 784). In the following Prague's lecture published on January 1921 \cite{Ein-GeomExp}, entitled \textit{Geometry and Experience} and translated in Einglish in \cite{Einstein-papers7bis}, he discussed on the applicability of GR to microscopic phenomena domain and concluded: `According to the view advocated here, the question whether this continuum [space-time] has a Euclidean, Riemannian, or any other structure is a question of physics proper which must be answered by experience, and not a question of a convention to be chosen on grounds of mere expediency.' (\cite{Einstein-papers7bis}; p. 214).
But he emphasized: `It is true that this proposed physical interpretation of geometry breaks down when applied immediately to spaces of sub-molecular order of magnitude.' (\cite{Einstein-papers7bis}; p. 214).

\subsection{Reconciling GR with microscopic phenomena}\label{new-idea}
In the same period, in December 1920, Einstein's friend Michele Besso reawakened his interest in Weyl's theory. Besso wrote Einstein wondering under which transformations Weyl's theory is invariant (\cite{Einstein-papers10bis}; p. 540). From this moment, Weyl's conformal transformations started to make their way in Einstein's mind a little at a time, as we shall see in the following. At the beginning of 1921 Einstein was invited to give lectures on the theory of relativity also in Vienna. During his lecture, Einstein should have talked about conformal transformations, because after returning to Berlin he corresponded with Wilhelm Wirtinger on this topic. In February, the Austrian mathematician wrote: `I pursued your remark in Vienna further, whether it would not be possible to form tensors that depended solely on the ratios of the $ g_{\mu\nu} $'s, and arrived at some quite satisfactory and interesting results.' (\cite{Einstein-papers12bis}; p. 44). In his first letter, Wirtinger analysed the effect on particle's paths obtained by varying an action principle using a new line element, `which also depends only on the ratios of the $ g_{\mu\nu} $'s.' (\cite{Einstein-papers12bis}; p. 45).  In his answer Einstein pointed out: `I am convinced that you have thus done relativity theory an inestimable service. For it is now a simple matter to construct a theory of relativity that assigns meaning to the ratios of the $ g_{\mu\nu} $'s, or the equation $ g_{\mu\nu}dx^\mu dx^\nu = 0 $, without--as with Weyl--, in my conviction, the physically meaningless quantities $ A_\mu $ (electromagnetic potentials) explicitly appearing in the equations. There only remains the problem of whether Nature really has made use of this possibility available to her to constrain herself accordingly. As soon as I have formed a judgement about this, I shall give myself the pleasure of informing you of the details.'\footnote{We changed the notation in order to make it uniform throughout our paper.} (\cite{Einstein-papers12bis}; p. 52) 
Hence, Einstein considered the possibility of introducing a sort of scale invariance in GR, in order to generalize it and to unify gravitational and electromagnetic phenomena. Einstein was interested in finding all conformally invariant tensors. 

In the same period Einstein wrote both to Hendrick A. Lorentz and to Ehrenfest with a great enthusiasm about his new attempt. February 22, to Lorentz: `I now have another hope of throwing a light \textit{on the realm of the molecular with the aid of relativity theory}. For there is a possibility that the following two postulates can be united with each other. 1) The natural laws depend only on the ratios of the $ g_{\mu\nu} $'s [...] [Weyl's postulate]. 2) The electromagnetic potentials do not explicitly enter into the laws, just the field strengths. I am very curious to see if these hypotheses will prove correct.' [emphasis added] (\cite{Einstein-papers12bis}; p. 51) On March 1, to Ehrenfest: `A good idea occurred to me about relativity. One can, like Weyl, assign physical meaning just to the relative value of the $ g_{\mu\nu} $'s ( i.e., to the light cone $ ds^2 = 0 $ ), \textit{without therefore having to resort to the characteristic} $ A $\textit{-metric with the non-integrable changes in the distances or measuring rods.}' [emphasis added] (\cite{Einstein-papers12bis}; p. 60). This statement clearly shows that Einstein considered the conformal invariance as a viable physical principle in order to extend the realm of GR to the microscopic scales.

As a result of his correspondence with Wirtinger, Einstein sent a contribution to the Prussian Academy: \textit{On natural addition to the foundation of general relativity}. In the paper, received on March 3, 1921, Einstein suggested considering a theory where only the ratios of the metric have physical meaning and emphasized that `[it] seems to me to be a lucky and natural one, even though one cannot a priori know whether or not it can lead to a useful physical theory.' (\cite{Einstein-papers7bis}; p. 225). He proposed also to study the particle's dynamics described by a new conformal invariant action\footnote{We will not go further into the details of the theory.}, constructed by referring explicitly to his correspondence with Wirtinger. He concluded his paper with the following emphasis. `Our only intention was to point out a logical possibility that is worthy of publication; it may be useful for physics or not. Only further investigations can show whether one or the other is the case [...]' (\cite{Einstein-papers7bis}; p. 228). 
Einstein's enthusiasm faded away soon: on June 30 Einstein wrote to Lorentz: `I also made an attempt at generalizing the theory but I am skeptical of it myself.' (\cite{Einstein-papers12bis}; p. 119). 
 
\subsection{Merging Weyl's and Kaluza's theory}\label{merge}
In May, 1921, Einstein gave four lectures at Princeton, where he only mentioned Weyl's and Kaluza's attempts: `A theory in which the gravitational field and the electromagnetic field do not enter as logically distinct structures would be much preferable. H. Weyl, and recently Th. Kaluza, have put forward ingenious ideas along this direction; but concerning them, I am convinced that they do not bring us nearer to the true solution of the fundamental problem. I shall not go into this further [...]' (\cite{Einstein-papers7bis}; p. 358)
About Weyl's principle, as we mentioned at the end of \ref{new-idea}, Einstein communicated his conviction also to Lorentz. About Kaluza's attempt, as we pointed out in section \ref{Kaluza}, he was aware of its incompatibility with electrons dynamics.

Even if Einstein maintained a sceptical approach, he continued to consider every possible development of these theories. In January, as we discussed in section \ref{Kaluza}, Einstein and Grommer sent their analysis on the existence of centrally symmetric solutions in Kaluza's theory. In 1922 he wrote to Weyl and wondered if he had read Kaluza's approach (\cite{Einstein-papers13bis}; p. 181). Einstein convinced himself step by step of the necessity of introducing further abstractions. In July 1923, he delivered a lecture to the Assembly of Nordic Naturalists in Gothenburg and declared: `The theory of gravitation--that is, viewed from the standpoint of mathematical formalism, Riemannian geometry--shall be generalized in such a way as to include the laws of the electromagnetic field. Unfortunately, in this endeavour, unlike the case of the derivation of the theory of gravitation (equivalence of inertial and gravitational mass), \textit{we cannot base our efforts on empirical facts}. Instead, \textit{we have to base them on the criterion of mathematical simplicity}, which is not free from arbitrariness.' [emphasis added] (\cite{Einstein-papers14bis}; p. 80).

Worried about the fact that GR was not able to explain microscopic phenomena, he continued to develop different attempts using new ideas. In his searching for connections between the theory of electron and GR he persevered to follow the unification way, always starting from the inclusion of the electromagnetic theory and GR in a wider framework. As pointed out by Daniela W\"unsch, `between 1919 and 1921 [...] he became convinced that this method of unification represented a significantly new path in physics' (\cite{Wuensch}; p. 277). She interpreted Einstein's approach for unification published in \textit{Nature} \cite{Einstein1923} as follows. In proposing an extension of the Eddington--Weyl theory at the beginning of 1923, `Einstein borrowed one element of Kaluza's theory: the Hamilton function contains a single tensor to describe the unified field.' (\cite{Wuensch}; p. 286). It is worth noting that Einstein did not mention explicitly Kaluza's approach in \cite{Einstein1923}, but he had already identified in \cite{EG-23}, as already pointed out at the beginning of section \ref{5-dim}, a unitary action as one of the advantages of Kaluza's approach. This means that W\"unsch inferred that Einstein tried to merge postulates of different approaches. Therefore, it is not so unreasonable to assert that in 1927 Einstein did it again and that he tried to merge Kaluza's attempt with Weyl's idea, because, as we argued in section \ref{Ein-27}, he considered conformally related metrics in five-dimensions, and hoped that this conformal transformations did not affect the four-dimensional physics. From this point of view, the $ \gamma_{55}=\text{constant} $ scenario appears as a gauge choice.

Both Einstein and Klein inserted their Ansatz into the action and varied it  considering only the $ g_{\mu\nu} $ and the $ A_\mu $ as Lagrangian variables. But Einstein presented this procedure as a choice of a particular reference system in five-dimensions, which would appear as a gauge fixing with respect of the conformal symmetry group in four dimensions. Therefore, Einstein's procedure is equivalent to the modern construction of a reduced action starting from a constrained action, i.e. an effective action, describing physical phenomena only up to a specific energy scale. As far as we know, the role of the cylinder conditions as a constraint condition for the action was pointed out for the first time in the Mid Seventies by Elhanan Leibovitz and Nathan Rosen \cite{Leibovitz-Rosen}. They proved that the Einstein-Maxwell field equations can be consistently derived by the five-dimensional curvature scalar, by imposing the cylinder and the sharpened cylinder conditions as constraints of the action. They also proved that this action can be further generalized. Einstein's idea of merging Weyl's and Kaluza's approach was reconsidered by Yu S. Vladimirov at the beginning of the Eighties. He recognized explicitly the connection between the transformations of fifth coordinate, namely $ \displaystyle (x^5)' = x^5 + \psi\left(\, \bbar{x^0},\, \bbar{x^1},\, \bbar{x^2},\, \bbar{x^3}\,\right) $, called by Vladimirov `special gauge transformations' (\cite{Vladimirov}; p. 1171), 
and the conformal transformations. Indeed, Vladimirov pointed out: ``The number of special gauge invariant 4-metrics forms a set of conformally corresponding [...] metrics' (\cite{Vladimirov}; p. 1172).

\section{Epilogue: really inconsistent?}\label{epilogue}
We started the Prologue by considering Klein's approach to five-dimensional relativity. We emphasized that Klein's approach was inconsistent from a fully five-dimensional point of view. The communications published by Einstein in 1927 are very similar to Klein's approach, in the sense that he did not obtained new results, as he himself declared. But Einstein's path into the five-dimensional world started many years before Klein, because of his role in Kaluza's early formulation of the theory. We raised many questions at the beginning of the paper: here we briefly summarize the answers we give in the precedent sections.

We tried to understand why Einstein published these two communications. Einstein was convinced that a unified theory should fulfil some precise requirements. One of them was that all the equations defining the model must have a covariant formulation. He struggled for years, from 1919 until 1927, with the non-covariant form of the cylinder condition. In the communications we analyzed, he finally exhibited a coordinate-free formulation. We emphasized that this was a reason for communicating them. From an historical perspective, with Einstein's paper the connection between Killing equation, space-time isometries and their link with electromagnetism gauge group emerged explicitly in the context of Kaluza-Klein theory. But he also discussed with Weyl the consequences of the introduction of a conformal structure. And in his attempts to construct a field theory able to shed light on the new microscopic phenomena, Einstein considered the conformal transformations as a viable theoretical principle. It was not unusual for Einstein to melt together different theoretical principles, struggling with the fact that he had no empirical facts to be guided by. In his first communications on the five-dimensional Universe, Einstein tried to merge Weyl's principle with Kaluza's ideas. Hence, this is a second reason for publishing his papers.

We found evidence, as far as we know for the first time, that Einstein could have inferred the inconsistency of Klein's model, because he had investigated the effect of $ \gamma_{55} $ on the four-dimensional physics and therefore he had calculated the full field equations, at least in the linear approximation, but he did not make any statement about it. Einstein knew Klein's work at that time, but their path and aims were different and we pointed out that there are some subtle differences between the two approaches. Unlike Klein, Einstein analysed the cylinder condition, the sharpened cylinder condition and the classical character of the fifth dimension more extensively\footnote{Klein was more interested in the quantum interpretation of the fifth dimension.}. 

Einstein wanted to unify electromagnetic and gravitational forces, but he wanted also to incorporate a deterministic description of the microscopic phenomena. Einstein viewed the unification path as a first step towards a full understanding of the microscopic phenomena. In 1921, he wrote to Lorentz that he hoped to extend the realm of GR to the microscopic phenomena with the help of Weyl's idea. Therefore, we are convinced that his introduction of the conformal transformations could reflect this attitude also in 1927. 

Einstein used a sort of modern gauge approach in order to write the action. As far as we know, this was the first time he used it in order to motivate his choice of the Lagrangian density, a point that Klein explicitly left out. Einstein suggested also to extend the gauge group of the electromagnetism to the group of conformal transformations in five dimensions which leave invariant the four-dimensional physics. Like Klein, Einstein decided to introduce the idea that only the ratios of the five-dimensional metric tensor components have physical meaning, but he was the father of this approach. Furthermore, Klein considered the possibility of giving physical meaning only to the ratios as a mere convention, while Einstein saw it as an important physical principle, which played a fundamental role in the construction of the Lagrangian density of his unified field theory.

We argued that by giving physical meaning only to the ratios of the five-dimensional metric, Einstein implicitly assumed that the fifth dimension had no physical significance in 1927. In the prologue, we emphasized that the inconsistency of Klein's model is connected with the fact that Klein tried to interpret the fifth dimension as real. Therefore, we can infer that in this sense Einstein's approach was not inconsistent, even if he introduced the sharpened cylinder condition. 

After this attempt, Einstein would change his attitude towards the fifth dimension in 1938 \cite{VanDongen} \cite{Witten}, but the ``road to reality'' for the extra-dimensions was not yet been completed \cite{Witten}.

\section*{Acknowledgement}
We express our gratitude to Morgan E. Aronson for invaluable comments and suggestions. We would like to acknowledge Jeroen van Dongen for sharing his opinion on Einstein's approach. This work has been supported in part by the DOR 2019 funds of the University of Padua and by the Smithsonian Libraries Dibner Library Resident Scholar Program, Washington DC.

\end{document}